# The concept «altruism» for sociological research: from conceptualization to operationalization

Oleg V. Pavenkov[a], Vladimir G. Pavenkov[b], Mariia V. Rubtcova[c1]

a. SPBGUKIT, 13, ul.Pravda, St.Petersburg, 191119, Russia
b. SPBGUKIT, 13, ul.Pravda, St.Petersburg, 191119, Russia
c. SPBGU, 7-9, Universitetskaya nab., St.Petersburg, 199034, Russia

**Abstract**:

This article addresses the question of the relevant conceptualization of «altruism» in Russian from the perspective sociological research operationalization. It investigates the spheres of social application of the word «altruism», include Russian equivalent «vzaimopomoshh`» (mutual help). The data for the study comes from Russian National Corpus (Russian). The theoretical framework consists of Paul F. Lazarsfeld`s Theory of Sociological Research Methodology and the Natural Semantic Metalanguage (NSM). Quantitative analysis shows features in the representation of altruism in Russian that sociologists need to know in the preparation of questionnaires, interview guides and analysis of transcripts.

**Resumen:**

En este artículo se aborda la cuestión de la conceptualización correspondiente de «altruismo» en ruso desde la perspectiva sociológica operacionalización investigación. Investiga las esferas de la aplicación social de la palabra «altruismo», incluye equivalente ruso «vzaimopomoshh`» (ayuda mutua). Los datos para el estudio proviene de Rusia National Corpus (en ruso). El marco teórico consta de Paul F. Lazarsfeld`s Teoría de Investigaciones Sociológicas Metodología y la semántica metalenguaje Natural (NSM). El análisis cuantitativo muestra características en la representación del altruismo en ruso que los sociólogos tienen que saber en la preparación de cuestionarios, guías de entrevistas y análisis de las transcripciones.

Palabras clave: Altruismo, categorías sociales, Natural Semantic metalenguaje (NSM), operacionalización sociológica, lingüística de corpus

[1]*Corresponding author. Tel.: +78122720003.
  *E-mail address:*m.rubtsova@spbu.ru

*1. Introduction*

This article presents a study of the conceptualization of «altruism» from the sociological operationalization perspective and includes small-scale study which analyzes how the concept of altruism and its Russian equivalent «vzaimopomoshh`» (mutual help) is used in Russian language. Operationalization is part of the empirical research process (Shields, Rangarjan 2013).

To translate the concept on "the language of the questionnaire", the development of in-depth semi-structured interview guide and to work with the transcript should be made operationalization - empirical interpretation of the concept, i.e. the procedure for describing concepts through a system of indicators, having the properties: a) measurable, b) the validity, c) repeatability. (Iliassov 2013)

The problem of definition of the spheres of application concepts is one of the main for sociological operationalization in empirical research. However it gained less development in application to Russian. Our research relies on researches of a Natural Semantic Metalanguage (NSM) (Wierzbicka 1972, 1992, 1997, 2002; Gladkova 2010, 2013). In opinion on this conception, linguist can compare different languages (and different terms into one language) objectively, if he (she) uses the semantic universals. However we are guided by discussions include the criticism of Natural Semantic Metalanguage (NSM) (Blumczyński 2013). In previous work, we have noted that the boundaries of social categories in the Russian language have difficulty in clear determining (Volchkova, Pavenkova, 2002; Rubtsova. 2007, 2011; Pavenkov 2014).

In our view, the study of the concept should include the study of synonyms, as sociologists and respondents are often used them. It can be said that the theme of the use of synonyms in sociological studies rarely studied. In Russian concepts altruism and «vzaimopomoshh`» (mutual help) are regarded as synonyms in many Russian theoretical concepts of Social Sciences. At the same time, they can be listed through a comma (Kropotkin 1988).

Therefore, our main research questions are following:

1) Are the concepts «altruism» and «vzaimopomoshh`» (mutual help) interchangeable?

2) Has the use of these words in oral and written speech essential differences?

Using the Russian National Corpus, we can establish similarities and differences in the use of word «altruism» in the context of «vzaimopomoshh`» (mutual help) as Russian equivalent and try to find answer on our research questions.

For this aim we can formulate and check folowing hypothesises:

*Hypothesis 1*. The using of these words has not statistically significant differences.

*Hypothesis 2*. The using of these words in oral and written speech has not statistically significant differences.

To confirm or disprove the hypotheses we describe the spheres of application these words. So we analyze seven contexts of using words «altruism» and «vzaimopomoshh`». They are people`s actions, human himself, relations` quality, state, social institutes, organisations, conseptions-ideology.

### *3. Data and Methodology*

Based on the fact that the Russian language has two concepts denoting the closest social phenomena – «altruism» and «vzaimopomoshh`» (mutual help) which consider as synonymous we give a characterization of their most frequent usage as a main, newspaper and spoken Russian National Corpus.

A corpus is a reference system based on an electronic collection of texts composed in a certain language. A national corpus represents that language at a stage (or several stages) of its development in all the variety of genres, styles, territorial and social variants of usage, etc. A national corpus is created by linguists for academic research and language teaching. A national corpus is distinguished by two features. Firstly, it is characterized by representative and well-balanced collections of texts. This means that such a corpus contains, if possible, all the types of written and oral texts present in the language (various genres of fiction, journalistic, academic, and business, as well as dialectal and sociolectal, texts). The proportion of text types in the corpus is based on their share in real-life usage at the time of composition. Secondly, a corpus contains additional information on the properties of texts that are included. This is achieved by means of annotation. The annotation is a principal feature of the corpus, distinguishing the corpus from simple collections (also known as 'libraries') of texts on the Internet. Such libraries are not well suited to academic work on the nature of language; they tend to focus on the content of texts rather than their language properties, while the creators of the Corpus recognize the importance of literary or scientific value of the texts, but see them as a secondary feature. Unlike an electronic library, the National Corpus is not a collection of texts which are deemed 'interesting' or 'useful' of themselves; the texts in the Corpus are interesting and useful for the study of language. Such texts might include not only great works of literature, but also works of a 'secondary' writer, or a transcription of an ordinary conversation[2].

|  | Corpus | | | |
|---|---|---|---|---|
|  | Total | Main | newspaper | Spoken |
| Total documents, sentences and words in Corpus |  | 85996 documents, 19362746 sentences, 229968798 words | 332720 documents, 12920590 sentences, 173518 798 words | 3525 documents, 1 623625 sentences, 10 754 403 Words |
| Word «altruism» in Corpus, including wordforms | 775 | 446 | 284 | 25 |
| Word «vzaimopomoshh`» (mutual help) in Corpus, including wordforms | 778 | 481 | 269 | 28 |
| Total words «altruism» and «vzaimopomoshh`» (mutual help) in Corpus, including wordforms | 1533 | 927 | 553 | 53 |

**Fig 1. Details of the words «altruism» and «vzaimopomoshh`» (mutual help) in Russian National Corpus**

The Russian National Corpus consists of (1533) relevant words, between (775) for altruism and (778) for «vzaimopomoshh`» (mutual help). The numbers of the words «altruism» and «vzaimopomoshh`» (mutual help) almost equally. At the same time distribution of the words is not equable. Main and spoken corpus of «vzaimopomoshh`» (mutual help) a little bit more than «altruism». Altruism is presented in newspaper lexis 5% more (Fig. 1). We have carried out check test in SPSS and confirmed that the two concepts are presented without statistical differences.

---
[2] Russian National Corpus http://www.ruscorpora.ru

| Al'truizm<br>Al'truist<br>Al'truisty<br>Al'truistichnyj<br>Al'truisticheskaja<br>Al'truistom<br>Al'truistka<br>Al'truistichnoj<br>Al'truisticheskom<br>Al'truizma | Altruism    Altruist<br>Altruists<br>altruistically<br>Altruistic<br>altruists<br>altruistka<br>altruistically<br>altruistic<br>Altruism | Vzaimopomoshh'<br>vzaimopomoshhi |

**Fig 2. Wordforms for concepts «altruism» and «vzaimopomoshh`-mutual help» in the Russian national corpus**

Also altruism and «vzaimopomoshh`» have the different wordforms in the Russian corpus (see Fig.2.)

## *4. Results and Analysis*

### 4. 1. The spheres of application «altruism» and «vzaimopomoshh`» (mutual help)

In order to show the differences in the spheres of application (context) of «altruism» and «vzaimopomoshh`-mutual help» (concepts) we will consider Figure (3) and Figure (4). As we can see the spheres of application of «altruism» and «vzaimopomoshh`-mutual help» have very big differences.

|  |  | The spheres of application – context |  |  |  |  |  |  | Total |
|---|---|---|---|---|---|---|---|---|---|
|  |  | people's actions | Human himself | relations' quality | state | social institutes | organisations | conceptions and ideology |  |
| Concepts | "vzaimopomoshh`" mutual help | 65 | 0 | 106 | 79 | 57 | 313 | 158 | **778** |
|  | Altruism | 180 | 98 | 188 | 14 | 10 | 15 | 250 | **755** |
| Total |  | 245 | 98 | 294 | 93 | 67 | 328 | 408 | **1533** |

**Fig 3. The spheres of application (context) of «altruism» and «vzaimopomoshh`-mutual help» (concepts)**

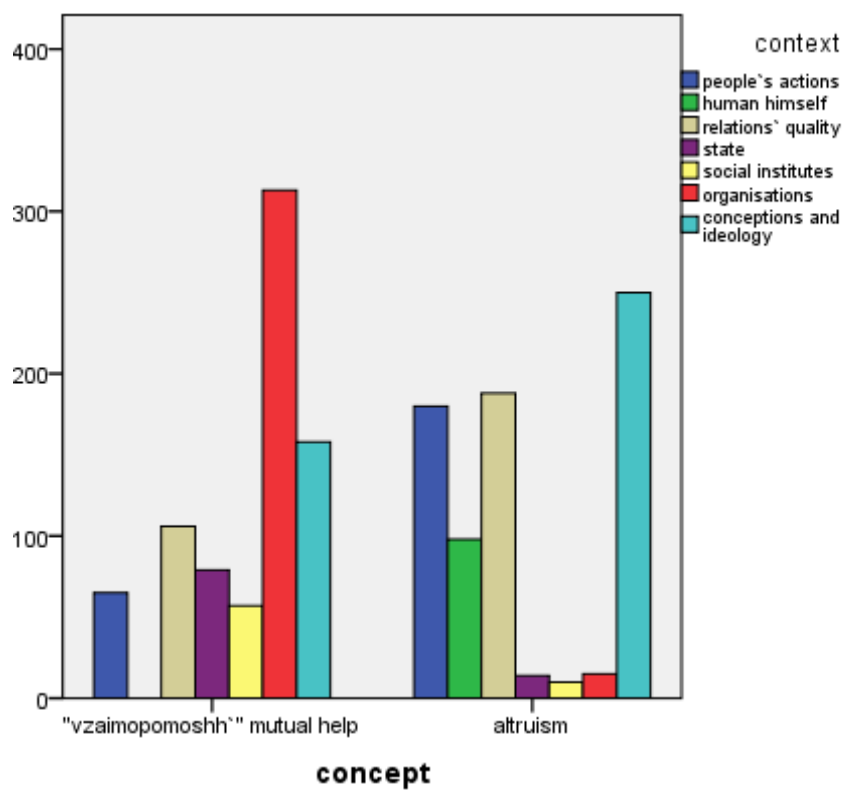

**Fig 4. The spheres of application (context) of «altruism» and «vzaimopomoshh`-mutual help»**

For statistical check we use Chi-Square Tests. Chi-Square Tests shows the significance of differences between the use of the word «altruism» and «vzaimopomoshh`-mutual help». ($p < 0,001$).Also, we confirmed the high level of significance of the differences in the use of the two words in a more rigorous test Cramer's V. ($p < 0,001$).

## 4. 2. The using of «altruism» and «vzaimopomoshh`» (mutual help) in oral and written speech: quantitative analyses

At the first step we have to find out is there a statistically significant difference between the use of concepts in oral and writing speech. We do not have a separate Russian written corpus. As the written corpus, we can assume the newspaper and the main corpuses, because they consist of written sources.

To check oral or written influence on the using of «altruism» and «vzaimopomoshh`» (mutual help) we have used ANOVA-test. It allows us to identify the factors influencing the use of concepts. The result is in the Fig 5.

**Tests of Between-Subjects Effects**

Dependent Variable: **concept**

| Source | | Type III Sum of Squares | df | Mean Square | F | Sig. |
|---|---|---|---|---|---|---|
| Intercept | Hypothesis | 364,731 | 1 | 364,731 | 1,470E3 | ,000 |
| | Error | 1,770 | 7,132 | ,248[a] | | |
| **Context** | Hypothesis | 44,377 | 6 | 7,396 | 20,216 | ,000 |
| | Error | 8,432 | 23,047 | ,366[b] | | |
| **Corpora** | Hypothesis | 1,024 | 2 | ,512 | ,892 | ,432 |
| | Error | 8,175 | 14,236 | ,574[c] | | |
| **context * corpora** | Hypothesis | 9,031 | 12 | ,753 | 4,791 | ,000 |
| | Error | 237,520 | 1512 | ,157[d] | | |

a. ,256 MS(corpora) + MS(Error)

b. ,351 MS(context * corpora) + MS(Error)

c. MS(context * corpora) + ,299 MS(Error)

d. MS(Error)

**Fig 5. Influence the choice of a linguistic corpus (main, newspaper or spoken) and context to represent the results on concepts «altruism» and «vzaimopomoshh`-mutual help»**

As we can see, the choice of a linguistic corpus (main, newspaper or speech) has not influence ($p = 0,432$) to represent the results on concepts «altruism» and «vzaimopomoshh`-mutual help». From another side, the context and interaction between context and corpus are strong statistically significant (Sig., $p < 0,001$). Chi-Square Tests shows the significance of differences between the use of the word «altruism» and «vzaimopomoshh`-mutual help» ($p < 0,001$) in main, newspaper or spoken linguistic corpuses. We can imagine them on a graph (Fig.6)

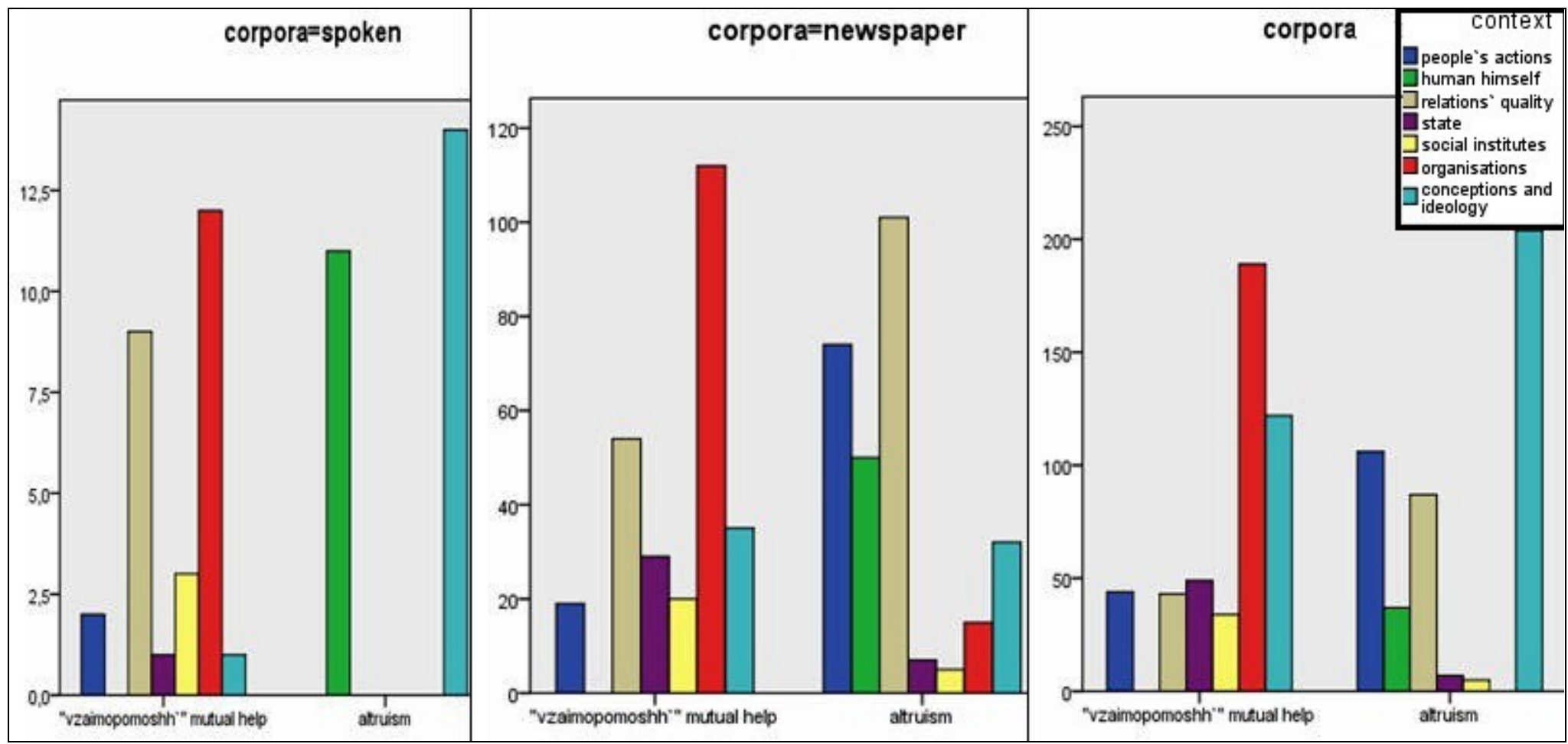

Fig. 5. The spheres of application (context) of «altruism» and «vzaimopomoshh`-mutual help» (concepts) in spoken, newspaper and main linguistic corpuses

*Discussion and conclusion*

We had explored the use of the two Russian words – «altruism» and «vzaimopomoshh`» (mutual help), which had the same meanings originally. The concepts «altruism» and «vzaimopomoshh`» (mutual help) are regarded as synonyms in many Russian theoretical Social Sciences` concepts. Based on Natural Semantic Metalanguage (NSM) and Russian National Corpus analysis we describe seven contexts of using words «altruism» and «vzaimopomoshh`»: people`s actions, human himself, relations` quality, state, social institutes, organisations, conseptions-ideology.

The quantitative analyse shows differences in the use of these concepts. The differences between the «vzaimopomoshh`» (mutual help) and «altruism» are statistically significant for all three Russian corpuses: main, spoken and newspaper. Only in few contexts these concepts may be used as synonyms: as features of quality of relations, as conception and ideology.

Concept «Altruism» is often combined with pronouns and it is more personal. The concept «vzaimopomoshh`» (mutual help) originally occurred from the life of the Russian local community. However, in the 20th century, the state has done its template. This explains its frequent mention in the context of organizations, social institutions and the state. In these patterns, the lyrical meaning of «happy life together» is almost lost.

In conclusion we would like to emphasize the importance of this operationalization of concepts for empirical sociological research. We agree with Lazarsfeld (Lazarsfeld, 1962:579) that the operationalization of concepts is required. The conceptualization of social categories and prevents the loss of meaning, which is important when we conduct in-depth interviews, focus groups and in data processing of transcripts.

In future studies, we would have paid more attention to the replenishment of the spoken Russian National Corpus with sociological interview transcripts, search for social differences in the use of terms and reconsidered these concepts.